\documentclass{article}
\usepackage{booktabs}
\usepackage{algorithm}
 \usepackage{algorithmic}
 \usepackage{multirow}
\usepackage{autobreak}
\usepackage{enumitem}
 \usepackage{caption}
 \usepackage{tabularx} 

\usepackage{PRIMEarxiv}

\usepackage[utf8]{inputenc} % allow utf-8 input
\usepackage[T1]{fontenc}    % use 8-bit T1 fonts
\usepackage{hyperref}       % hyperlinks
\usepackage{url}            % simple URL typesetting
\usepackage{booktabs}       % professional-quality tables
\usepackage{amsfonts}       % blackboard math symbols
\usepackage{nicefrac}       % compact symbols for 1/2, etc.
\usepackage{microtype}      % microtypography
\usepackage{lipsum}
\usepackage{fancyhdr}       % header
\usepackage{graphicx}       % graphics
\graphicspath{{media/}}     % organize your images and other figures under media/ folder

\title{Learning to Collaborate via Structures: Cluster-Guided Item Alignment for Federated Recommendation
}

\author{
  Yuchun Tu \\
  Beihang University \\
  Beijing \\
  China\\
  \texttt{tuyuchun@buaa.edu.cn} \\
   \And
  Zhiwei Li \\
  University of Technology Sydney \\
  Sydney \\
  Australia\\
  \texttt{zhw.li@outlook.com} \\
  \And
  Bingli Sun \\
  Beihang University \\
  Beijing \\
  China \\
  \texttt{youthbl@buaa.edu.cn} \\
  \And
  Yixuan Li \\
  University of Edinburgh \\
  Edinburgh \\
  UK \\
  \texttt{yixuan.li.cp@ed.ac.uk} \\
  \And
  Xiao Song \\
  Beihang University \\
  Beijing \\
  China \\
  \texttt{songxiao@buaa.edu.cn} \\
}

\begin{document}
\maketitle

\begin{abstract}
Federated recommendation facilitates collaborative model training across distributed clients while keeping sensitive user interaction data local. Conventional approaches typically rely on synchronizing high-dimensional item representations between the server and clients. This paradigm implicitly assumes that precise geometric alignment of embedding coordinates is necessary for collaboration across clients. We posit that establishing relative semantic relationships among items is more effective than enforcing shared representations. Specifically, global semantic relations serve as structural constraints for items. Within these constraints, the framework allows item representations to vary locally on each client, which flexibility enables the model to capture fine-grained user personalization while maintaining global consistency. To this end, we propose \textbf{C}luster-\textbf{G}uided \textbf{FedRec} framework (CGFedRec), a framework that transforms uploaded embeddings into compact cluster labels. In this framework, the server functions as a global structure discoverer to learn item clusters and distributes only the resulting labels. This mechanism explicitly cuts off the downstream transmission of item embeddings, relieving clients from maintaining global shared item embeddings. Consequently, CGFedRec achieves the effective injection of global collaborative signals into local item representations without transmitting full embeddings. Extensive experiments demonstrate that our approach significantly improves communication efficiency while maintaining superior recommendation accuracy across multiple datasets.
\end{abstract}

% keywords can be removed

\section{Introduction}
Recommendation systems are fundamental to modern online platforms for delivering personalized content. 
However, traditional centralized recommendation paradigms require collecting massive user behavior data to a central server, raising significant privacy concerns and violating strict data regulations such as GDPR \cite{protection2018general} and CCPA \cite{yang2019federated}. 
To address these issues, Federated Learning (FL) has emerged as a privacy-preserving alternative \cite{konevcny2016federated}. 
Federated Recommendation (FedRec) Systems integrate FL with recommendation algorithms, which allow distributed clients to collaboratively train models while keeping raw user interaction data local \cite{koren2009matrix,sun2024survey,jiangTutorialPersonalizedFederated2024}. 
Despite its privacy benefits, FedRec faces critical challenges in communication efficiency and representation robustness.

Most existing FedRec methods follow a parameter-aggregation paradigm. 
The central server aggregates and broadcasts high dimensional item embeddings to align user preferences across clients \cite{luoPersonalizedFederatedRecommendation2022,zhangDualPersonalizationFederated2023}. 
It implicitly assumes that precise geometric alignment of embedding coordinates is necessary for global collaboration. 
However, this assumption leads to two major limitations. 
First, transmitting dense embedding matrices incurs prohibitive communication costs. 
The overhead scales linearly with both the number of items and the embedding dimension, creating a bandwidth bottleneck for resource-constrained edge devices~\cite{li2024federated}. 
Second, strict coordinate alignment is often ineffective under non-IID data distributions. 
Client-specific data heterogeneity causes significant drift in local representation spaces \cite{han2025fedcia}. 
Directly averaging divergent embeddings results in noisy global representations and degrades recommendation accuracy.

We argue that effective collaboration does not require sharing exact embedding coordinates. 
Instead, it relies on sharing the underlying semantic structure of items, as the topological relationship among items is more robust than their absolute geometric positions~\cite{liu2024deeply}. 
Specifically, the intrinsic semantic proximity among items acts as a coordinate-invariant property that remains consistent across heterogeneous client distributions. 
Therefore, the server should function as a global structure discoverer rather than a simple parameter aggregator. 
Transmitting discrete structural signals can decouple collaborative knowledge from high-dimensional parameters. 
This perspective offers a pathway to break the dependency between communication cost and embedding size in FedRecs.

\begin{figure}
\centerline{\includegraphics[width=\linewidth]{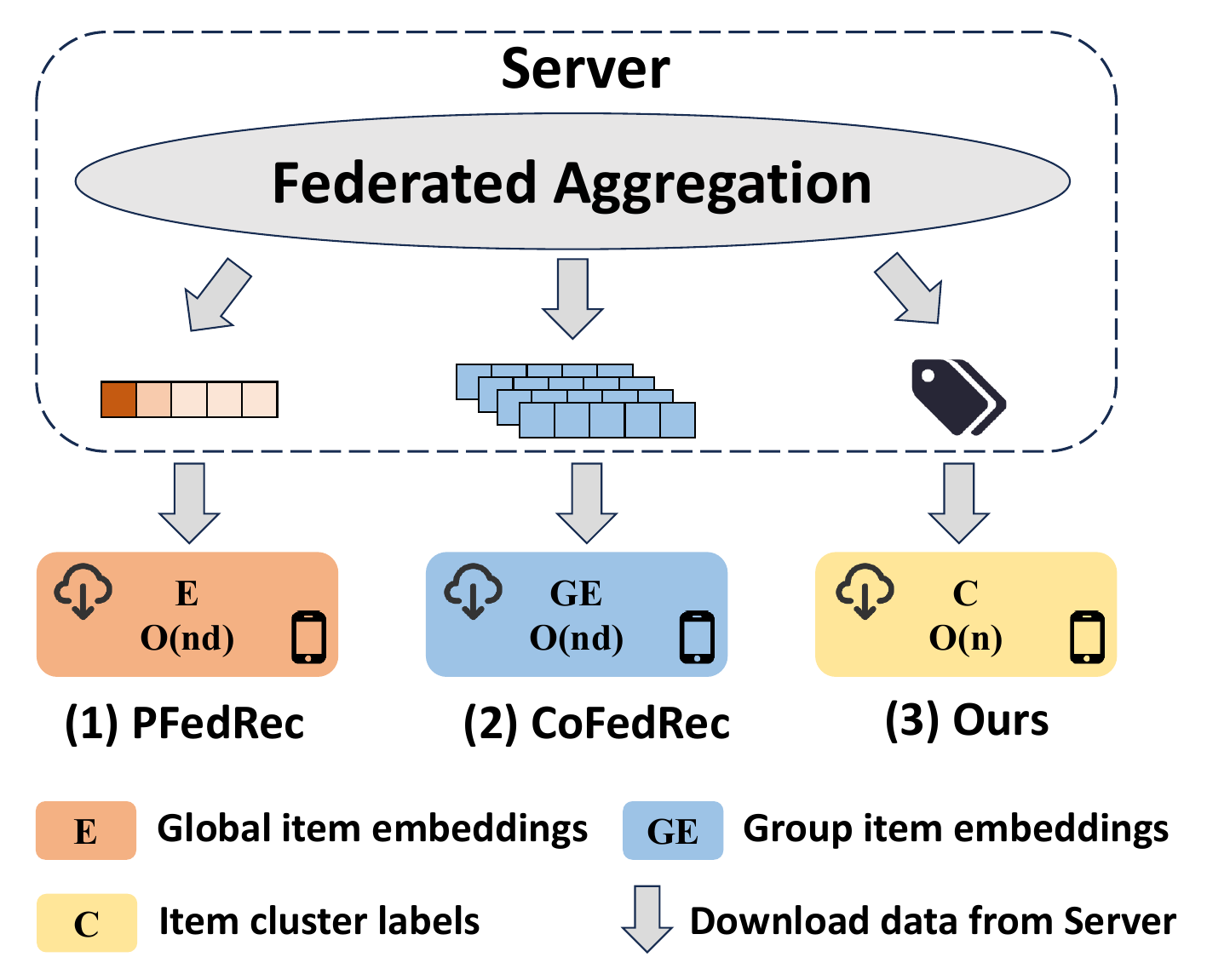}}
\caption{Comparison of representative federated recommendation paradigms for communication. $O(\cdot)$ denotes the communication complexity.}
\label{paradigms}
\end{figure}

Motivated by this insight, we propose a \textbf{C}luster-\textbf{G}uided \textbf{FedRec} framework (CGFedRec).
We reformulate the federated update process as a structure learning task.
The server aggregates client updates to discover global item clusters, and then broadcasts only compact cluster labels instead of dense embeddings.
These labels serve as lightweight structural constraints.
On the client side, we design a structure-aware alignment mechanism, where clients utilize the received labels to align their local item representations with such a global semantic structure.
This allows clients to maintain personalized embedding spaces while adhering to global consistency.
Consequently, CGFedRec explicitly decouples communication overhead from representation capacity, reducing the transmission complexity to be independent of the embedding dimension.

In summary, the main contributions of this work are as follows: \begin{itemize}[left=0pt] 
\item 
We propose a structural collaboration perspective for FedRec to demonstrate that aligning global semantic structures is more effective and efficient than synchronizing embedding coordinates. 

\item 
We present CGFedRec, a novel framework that replaces embedding transmission with cluster label sharing, explicitly decoupling communication overhead from model dimensionality. 

\item 
We develop a cluster-guided contrastive alignment strategy for FedRec that effectively mitigates data heterogeneity by injecting global structural constraints into local personalized training. 

\item Extensive experiments on multiple real-world datasets show that CGFedRec outperforms state-of-the-art methods in both recommendation accuracy and communication efficiency. 
\end{itemize}
\section{Related Work}

\subsection{Federated Recommendation Systems}
Federated recommendation enables multiple clients to jointly train recommender models without uploading raw user-item interaction data\cite{ammad2019federated, sun2024survey,wangPersonalizedFederatedContrastive2025,wangFederatedRecommendationExplicitly2025,heCoclusteringFederatedRecommender2024}. This paradigm reduces privacy risks and supports collaborative learning under data decentralization. Early studies mainly extend traditional recommendation models into the federated setting\cite{ammad2019federated,perifanisFederatedNeuralCollaborative2022,chaiSecureFederatedMatrix2021,wuFederatedGraphNeural2022}. Recent work further considers user heterogeneity, where clients have non-IID behaviors and personalized preferences. PFedRec \cite{zhangDualPersonalizationFederated2023} introduces dual personalization to generate user-specific views for item representations. FPFR \cite{wangFairPersonalizedFederated2024} integrates graph neural networks, user clustering, and fairness constraints to balance recommendation accuracy and group fairness. FedDAE\cite{li2025personalized} combines variational autoencoders with dual encoders (global/local) and adaptive gating to achieve personalized recommendations. Although effective, these methods still depend on synchronizing item embeddings or model parameters, which incurs communication overhead that scales with the item size and embedding dimension and becomes a bottleneck in large-scale settings. Existing methods reduce communication via compression techniques such as low-rank factorization and quantization\cite{han2025fedcia,chen2025beyond}, but they still transmit embeddings or parameters, which remain costly for large item spaces. This limitation motivates alternative collaboration paradigms that share compact semantic structures rather than full embeddings. Instead of enforcing strict representation-level alignment, structural information can serve as global constraints, while local item representations remain flexible for personalization. Such a direction provides a foundation for communication-efficient federated recommendation frameworks.

\subsection{Clustering Federated Learning}
Federated Learning (FL) is a distributed learning paradigm that enables multiple clients to collaboratively train a shared model without sharing their local data \cite{li2020federatedfuturedirections, konevcny2016federated, mcmahan2017communication, 9599369}. Nevertheless, the presence of non-IID client data often results in significant performance degradation in federated models, posing challenges to meeting personalized requirements in real-world applications \cite{li2020federateoptimization, zhao2018federated}. To address the challenges posed by statistical heterogeneity, various strategies have been proposed, including personalized modeling \cite{caiFedCEPersonalizedFederated2023, liangEfficientOneoffClustering2023}, meta-learning \cite{liu2023federated, fallah2020personalized}, and clustering methods \cite{yanClusteredFederatedLearning2024, panMachineUnlearningFederated2023, ruanFedSoftSoftClustered2022, qiaoFederatedSpectralClustering2023, duDynamicAdaptiveIterative2023, kimDynamicClusteringFederated2021}. Clustered federated learning groups clients with similar data distributions and shares models within each group, thereby enhancing local adaptability while maintaining global generalization. This makes it an effective approach for addressing non-IID challenges. \cite{ghoshEfficientFrameworkClustered2020, longMulticenterFederatedLearning2023, sattlerClusteredFederatedLearning2021}. FedCDC \cite{sun2025fedcdc} uses graph-based community detection (Louvain algorithm) to dynamically cluster clients by similarity without predefined thresholds, achieving superior accuracy on non-IID data. Furthermore, clustering federated learning incorporates clustering techniques into FedRec to tackle the heterogeneity of user preference distributions \cite{yeAdaptiveClusteringBased2024, heCoclusteringFederatedRecommender2024, maoClusterdrivenPersonalizedFederated2024}. ClusterFedMeta \cite{yuFederatedRecommendationAlgorithm2024} utilizes user clustering and meta-learning to address challenges posed by non-IID data, including low communication efficiency and insufficient model personalization.

\section{Methodology}
\subsection{Overall Framework}
\begin{figure}
    \centering
    \includegraphics[width=\linewidth]{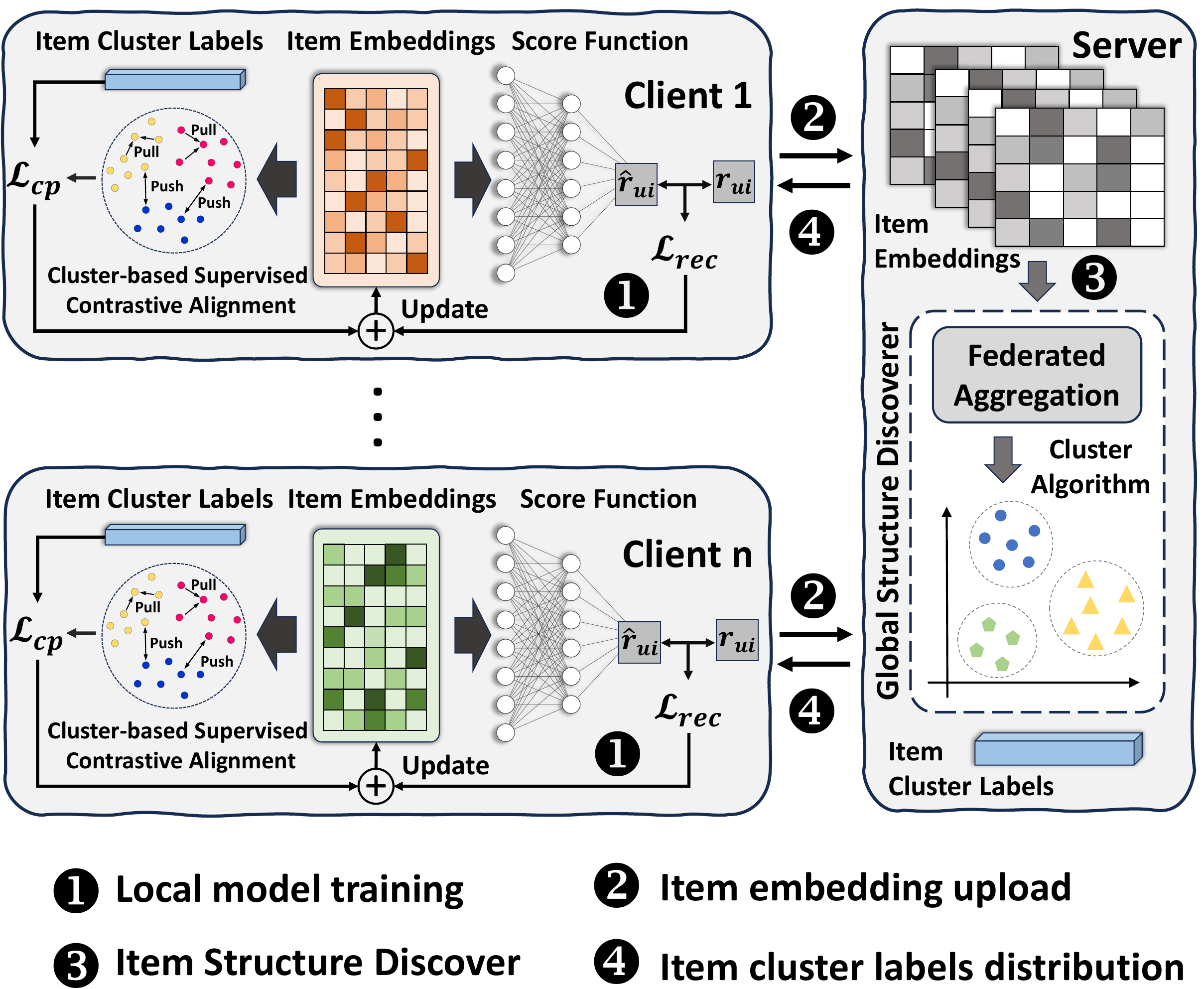}
    \caption{The framework of CGFedRec. Each client first trains a local recommendation model to learn item representations. The server aggregates the uploaded representations and discovers global item structures via clustering, producing cluster labels for all items. Instead of broadcasting global embeddings, the server only sends these lightweight cluster labels back to clients. Clients then use the labels as structural supervision to align item representations through contrastive learning, enabling global knowledge injection with significantly reduced communication cost.}
    \label{fig:framework}
\end{figure}

In this section, we propose CGFedRec, a cluster-guided federated recommendation framework that communicates global semantic structures instead of synchronizing high-dimensional item embeddings, the framework is shown in Figure \ref{fig:framework}. 

% Different from conventional FedRec methods that enforce strict coordinate alignment of item representations, CGFedRec treats item clusters as the primary collaborative signal. The server learns a global clustering structure from uploaded client embeddings and broadcasts only the resulting cluster labels, which serve as structural constraints for local training.

% On each client, cluster labels are incorporated into a cluster-based supervised contrastive objective, where items within the same cluster are treated as positives and items from different clusters are treated as negatives. This design aligns the relative semantic relations among items while allowing local embedding spaces to remain flexible, thus improving robustness under non-IID user behaviors.
% cluster-level signals instead of global item embeddings to guide the training process, thereby mitigating noise introduction.
\subsection{Problem Formulation}
FedRec aims to enable personalized recommendation services while preserving user privacy by collaboratively training the recommendation model across multiple clients. 

\textbf{Notations.} Assume there are $n$ users (or clients), denoted by $\mathcal{U}=\{u_1, u_2, ..., u_n \}$, and $m$ items, denoted by $\mathcal{I}=\{i_1, i_2, ..., i_m\}$. Each user $u_p$ interacts with the item set $\mathcal{I}$ to generate local data $D_p$, where each data instance is represented as a triplet $\left(u_p, i, r_{p,i}\right)$, with $u_p \in \mathcal{U}$, $i\in \mathcal{I}$, and $r_{p,i} \in \left[0,1\right]$ indicating the interaction preference of user $u_p$ on item $i$ (e.g., click, view, or rating). In the proposed framework, the optimal objective is to minimize the empirical risk functions across all clients
\begin{equation}
    \mathop{\arg\min}\limits_{\left(E;\Theta_1,\Theta_2,...,\Theta_n\right)}\frac{1}{n}\sum_{p=1}^{n}\alpha_{p} \mathcal{L}_{p}\left(E;\Theta_{p};\mathcal{D}_{p}\right),
\end{equation}
where $\mathcal{L}_{p}\left(E;\Theta_{p};\mathcal{D}_{p}\right)$ denotes the loss function of the recommendation model on the $p$-th client's local dataset $\mathcal{D}_p$. The matrix $E \in \mathbb{R}^{m \times d}$ denotes the global item embeddings, where $d$ is the embedding dimension, and $\Theta_p$ denotes the local model parameters. The coefficient $\alpha_{p}=\frac{|\mathcal{D}_p|}{|\mathcal{D}|}$ specifies the aggregation weight, which is proportional to the relative size of client $p$’s dataset.

\subsection{Recommendation Model}
Client $p\in\mathcal{P}$ maintains an item embedding table $E_p \in\mathbb{R}^{m \times d} $. For an item $i$, its embedding is denoted as $e_{p,i}\in\mathbb{R}^{d}$. Following the implementation, we adopt an MLP-based predictor with a scoring head. The predicted preference score is computed as:

\begin{equation}
    \hat{r}_{p,i}=\sigma\left(w^{\top} e_{p,i}\right),
\end{equation}
where $w$ is the linear weight parameter and $\sigma(\cdot)$ denotes the sigmoid function. 
The recommendation loss on client $p$ employs binary cross-entropy, which defined as follows:
\begin{equation}
    \mathcal{L}_{p}^{r e c}=-\sum_{(p,i)\in D^+}\log\hat{r}_{p,i}-\sum_{(p,j)\in D^-}\log(1-\hat{r}_{p,j}),
\end{equation}
where $D^+$ denotes denotes the set of positive user pairs $(p,i)$, meaning that client $p$ has interacted with item $i$ (i.e., an observed interaction exists between them), and $D^-$ denotes the set of negative user pairs $(p,i)$, meaning that client $p$ has no interacted with item $j$.

\subsection{Cluster-Guided Structural Signal}
Our key design is to represent global collaborative knowledge as discrete semantic structures. At each communication round $t$, client $p$ uploads its interacted item embeddings $E^t_p$ to the server. The server then aggregates them to obtain global item embeddings:

\begin{equation}
    E^t_g=\frac{1}{|\mathcal{P}_t|}\sum_{p\in\mathcal{P}_t}E^t_p
\end{equation}
Instead of broadcasting $E^t_g$ back to clients, the server performs K-means clustering over item embeddings:
\begin{equation}
    \mathcal{C}^t=kmeans(E^t_g, k),
\end{equation}
where $k$ is the number of clusters. Each item $i$ is assigned a discrete cluster label $c^t_i\in\{1,2,\cdots, k\}$. The server then broadcasts only the cluster assignment vector:
\begin{equation}
    \mathcal{C}^t=\{c^t_i\}_{i\in \mathcal{I}}
\end{equation}
% which reduces the downstream communication complexity from $O(|\mathcal{I}|d)$ to $O(\mathcal{I})$.

\subsection{Cluster-based Structural Alignment}
To inject the global structure into local representation learning, each client constructs a supervised contrastive loss based on the received cluster labels. Let $z_{p,i}$ denote the normalized embedding of item $i$ on client $p$:
\begin{equation}
    z_{p,i} = \frac{e_{p,i}}{||e_{p,i}||_2}.
\end{equation}
For each item pairs $(i,j)$, we define its positive set as:
\begin{equation}
    \Phi(i,j) = \{(i,j)| i\neq j,c_i=c_j \},
\end{equation}
and all remaining items are treated as negatives.

To implement this contrastive learning mechanism, we construct a mask matrix $\mathcal{M} \in \mathbb{R}^{m \times m}$, where the entries corresponding to items at the same cluster are set to 1, and set otherwise to 0:

\begin{equation}
    \mathcal{M}_{ij}= 
    \begin{cases} 
        1 & \mathrm{if~}(i,j) \in \Phi(i,j)\\
        0 & \mathrm{otherwise}
    \end{cases}.
\end{equation}

For each item $i$ on client $p$, we then calculate the mean log-probability over all positive sample pairs within the same cluster:
\begin{equation}
    \Omega_p(i)=\frac{1}{\sum_{j}\mathcal{M}_{ij}}
\sum_{j=1}^{m}\mathcal{M}_{ij} \cdot \log
\frac{
  \exp\!\bigl(e_{p,i} \cdot e_{p,j}^\top/ \tau\bigr)
}{
  \displaystyle \sum_{l \neq i}^m \exp\!\bigl(e_{p,i} \cdot  e_{p,l}^\top / \tau\bigr)
}
\label{eq:temp}
\end{equation}
where $e_i$ is the embedding of item $i$, $\tau$ is a temperature parameter that controls the smoothness of the distribution. 
% The temperature parameter $\tau$ serves as a distribution sharpening regulator that fundamentally controls the concentration of the softmax distribution over positive and negative sample pairs. 
% Specifically, smaller $\tau$ values induce sharper distributions that emphasize hard negative mining, promoting tighter intra-class clustering and enhanced inter-class separability, while larger $\tau$ values produce smoother distributions that facilitate stable optimization but may compromise discriminative capacity, with adaptive calibration based on dataset characteristics. 
Finally, we calculate the average supervised contrastive loss over all items as:
\begin{equation}
\label{eq:sc_loss}
\mathcal{L}_{cg}
= -\,\frac{\tau}{\tau_{\mathrm{base}}}\,\frac{1}{m}
\sum_{i=1}^{m} \Omega_p(i).
\end{equation}

\subsection{Overall Objective}

\begin{algorithm}[tb]
\caption{CGFedRec framework.}
\label{alg:overwork}

\begin{tabular}{@{}l@{\hspace{3pt}}p{0.85\linewidth}@{}}
\textbf{Input:}  &
Initialize global item embeddings $E_{0}$, participation rate $\gamma$, all participants $\mathcal{P}$, 
each holds a set of training data $\mathcal{D}$, cluster label of items $\mathcal{C}$, number of clusters $k$ \\

\textbf{Output:} &
Each client's item embeddings $E$, model parameters $\Theta$
\end{tabular}

\begin{algorithmic}[1]
\FOR{ each round $t=1, 2, \ldots , T$}
    \STATE $\mathcal{P}_t$ $\leftarrow $Randomly select subset of $ \gamma * \left| \mathcal{P} \right|$
    \FOR{ client $p\in\mathcal{P}_{t}$ in parallel}
        \STATE \# Client Update
        \STATE Download the cluster labels $\mathcal{C}$ from server
        \STATE Calculate $\mathcal{L}_p$ with Eq.(\ref{loss});
        \STATE $\Theta^{t}_p\leftarrow\Theta^{t-1}_p-\eta\nabla_{\Theta^{t-1}_p}\mathcal{L}_p$;
        \STATE Calculate $\mathcal{L}_p$ with Eq.(\ref{loss});
        \STATE $E^{t}_p \leftarrow E^{t-1}_p-\eta^{\prime}\nabla_{E^{t-1}_p}\mathcal{L}_p$
        \STATE Send the updated item embeddings $E_{p}^t$ to server
    \ENDFOR
    \STATE \# Global aggregation;
    \STATE $E_{g}^t \leftarrow \frac{1}{|\mathcal{P}_t|}\sum_{p=1}^{|\mathcal{P}_t|}E_{p}^t$;
    \STATE $\mathcal{C}^t \leftarrow Kmeans(E_{g}^t, k)$
    \STATE Send the cluster labels $\mathcal{C}^t$ to all clients
\ENDFOR
\end{algorithmic}
\end{algorithm}
Each client jointly optimizes the recommendation loss and the cluster-guided contrastive loss:
\begin{equation}
\label{loss}
    \mathcal{L}_p=\mathcal{L}_p^{rec}+\lambda\mathcal{L}_p^{cg}
\end{equation}

This objective enforces that items belonging to the same cluster are pulled closer, while items from different clusters are pushed apart. Importantly, it does not require coordinate-level consistency across clients, but only preserves relative semantic relations. The procedure is detailed in Algorithm \ref{alg:overwork}.

\subsection{Optimization and Communication Protocol}

CGFedRec follows a round-based federated optimization protocol. At round $t$, each participating client $p\in\mathcal{P}_t$ receives cluster labels $\mathcal{C}^t$ from the server. Then it performs local gradient descent updates:
\begin{equation}
\begin{aligned}
    \Theta_{p}^{t}\leftarrow\Theta_{p}^{t-1}-\eta\nabla_{\Theta_{p}}\mathcal{L}_{p},\\
    E_{p}^{t}\leftarrow E_{i}^{p-1}-\eta^{\prime}\nabla_{E_{p}}\mathcal{L}_{p},
\end{aligned}
\end{equation}
where $\eta$ and $\eta^{\prime}$ denote learning rates for recommendation model parameters and item embeddings, respectively.

After local training, the client uploads its updated item embeddings $E^t_p$. The server aggregates item embeddings to compute $E^t_g$, performs K-means clustering, and broadcasts the updated cluster labels for the next round.

\subsection{Communication Efficiency}
In this section, we compare the communication complexity of embedding synchronization-based FedRec methods with our clustering-label-based CGFedRec. Clearly, transmitting embeddings incurs a communication cost of $O(nd)$, where $n$ denotes the number of items (or users) and $d$ is the embedding dimension, while transmitting clustering labels only requires 
$O(n)$. Moreover, in practical scenarios, the communication gap is further enlarged because embeddings are typically represented by floating-point numbers, whereas clustering labels are stored as integers.

Specifically, in each communication round, traditional methods require the server to distribute full item global embeddings to every client, resulting in a high communication cost as follows:
\begin{equation}
    T_{base} = |\mathcal{P}_t| \times m \times d \times s_{f}
\end{equation}
where $s_{f}$ denotes the byte size of each floating-point number (typically 4 bytes). Since CGFedRec only transmits the cluster labels of each item, the total download data volume is:
\begin{equation}
    T_{ours} = |\mathcal{P}_t| \times m \times s_{i}
\end{equation}
where $s_{i}$ denotes the byte size of each integer number (typically 1 bytes). Thus, our communication saving rate is:
\begin{equation}
\label{eq: reduction}
    Reduction = 1-\frac{T_{ours}}{T_{base}} = 1-\frac{s_{i}}{d \times s_{f}}
\end{equation}
From (\ref{eq: reduction}), it is evident that varying latent dimensions lead to differences in computational cost.

Traditional approaches to enhancing model expressiveness (increasing dimension $d$) entail a linear communication cost. This constrains the capacity of federated recommendation systems to utilise large language models. CGFedRec breaks this dependency. This enables the use of extremely high-dimensional embedding vectors to capture fine-grained features without imposing any additional downstream communication burden. This characteristic holds significant appeal for practical industrial applications.

\section{Experiment}
\begin{table}[htbp]
\centering
\label{tab:statistics}
\renewcommand{\arraystretch}{1.2}
% \resizebox{\linewidth}{!}{
\begin{tabular}{@{}lrrrc@{}}
\toprule
\textbf{Datasets} & \textbf{\#Users} & \textbf{\#Items} & \textbf{\#Interactions} & \textbf{Sparsity (\%)} \\
\midrule 
\textbf{MovieLens-100K} & 943 & 1,682 & 100,000 & 93.70 \\
\textbf{MovieLens-1M}   & 6,040 & 3,706 & 1,000,209 & 95.53 \\
\textbf{KU} &204 &560 &3,489 &96.95\\
\textbf{Beauty} &253 &356 &2,535 &97.19\\
\textbf{FilmTrust}     & 1,227 & 2,059 & 34,888 & 98.62 \\ 
% \textbf{Supplier} &1081 &21517 & & \\
\bottomrule
\end{tabular}
\caption{Dataset Statistics. We selected 5 datasets with varying degrees of sparsity to validate our model's performance and show their statistical characteristics.}
\label{tab:datasets}
\end{table}
\begin{table*}[t]
\centering

\label{tab:main_results}
\renewcommand{\arraystretch}{1.2}
\resizebox{\textwidth}{!}{
\begin{tabular}{lcccccccccc}
\toprule
\multirow{2}{*}{\textbf{Method}} 
& \multicolumn{2}{c}{\textbf{ML-100K}} 
& \multicolumn{2}{c}{\textbf{ML-1M}} 
& \multicolumn{2}{c}{\textbf{Filmtrust}} 
& \multicolumn{2}{c}{\textbf{KU}} 
& \multicolumn{2}{c}{\textbf{Beauty}} \\

& \textbf{HR@5} & \textbf{NDCG@5} 
& \textbf{HR@5} & \textbf{NDCG@5}
& \textbf{HR@5} & \textbf{NDCG@5}
& \textbf{HR@5} & \textbf{NDCG@5}
& \textbf{HR@5} & \textbf{NDCG@5} \\
\midrule

\textbf{PFedRec}   & 0.5239 & 0.3684 & 0.5475 & 0.3829 & 0.9006 & 0.8170 & 0.3676 & 0.2826 & 0.0632 & 0.0355 \\
\textbf{GPFedRec}  & 0.4867 & 0.3342 & 0.4308 & 0.2668 & 0.8900 & 0.8105 & 0.2696 & 0.2079 & 0.0514 & 0.0287 \\
\textbf{CoFedRec}  & 0.4613 & 0.3539 & 0.4877 & 0.3721 & \underline{0.9234} & \underline{0.8417} & 0.2206 & 0.1522 & 0.1265 & 0.0803 \\
\textbf{FedRAP}    & 0.4602 & 0.3748 & 0.3162 & 0.2747 & 0.5100 & 0.4189 & 0.4783 & 0.4565 & 0.6173 & 0.5828 \\
\textbf{FedCA}     & \underline{0.7975} & \underline{0.7300} & \underline{0.7571} & \underline{0.6714} & 0.8862 & 0.8048 & \underline{0.7391} &\textbf{0.6483} & \underline{0.9259} & \underline{0.8355} \\

\textbf{CGFedRec}  & \textbf{0.9777} & \textbf{0.8927} & \textbf{0.9533} & \textbf{0.8851} & \textbf{0.9348} & \textbf{0.8512} & \textbf{0.7402} & \underline{0.5089} & \textbf{0.9565} & \textbf{0.8458} \\
\midrule
\textbf{Improvement} &$\uparrow22.60\%$ &$\uparrow22.29\%$ &$\uparrow25.91\%$ &$\uparrow31.83\%$ &$\uparrow1.23\%$ &$\uparrow1.13\%$ &$\uparrow0.15\%$ &$\downarrow21.50\%$ &$\uparrow3.30\%$ & $\uparrow1.23\%$\\
\bottomrule
\end{tabular}}
\caption{Performance comparison of different FedRec methods on five benchmark datasets in terms of HR@5 and NDCG@5. The best result for each dataset-metric pair is highlighted in "\textbf{bold}", while the second-best result is marked with "\underline{underline}". The "Improvement" row reports the relative percentage gain of the best-performing method over the second-best method.}

\label{tab:performance_comparison}
\end{table*}

To comprehensively evaluate the proposed method, we design experiments to answer the following research questions:
\begin{itemize}[left=0pt]
  \item \textbf{Q1:} Does CGFedRec outperform state-of-the-art federated recommendation methods?
  \item \textbf{Q2:} How does our proposed Cluster-Guided federated recommendation method work?
  \item \textbf{Q3:} Can CGFedRec preserve personalized item embeddings while achieving globally consistent semantic structures?
  \item \textbf{Q4:} How do the key hyperparameters of CGFedRec affect its performance?
\end{itemize}

\subsection{Datasets and Evaluation}

 \textit{\textbf{Datasets.}}
To comprehensively evaluate the performance of the proposed CGFedRec framework, we conduct experiments on 5 different sparsity public datasets: ML-100k, ML-1M \cite{harper2015movielens}, KU\cite{zhang2024ninerec}, Beauty\cite{hou2024bridging}, and FilmTrust\cite{guo2013novel}. According to \cite{he2020lightgcn}, we retain user records that contained more than 5 interactions during pre-processing. Table \ref{tab:datasets} presents the statistics of the processed datasets. ML-100k and ML-1M record users' preference ratings on movies over time. KU dataset contains users’ interaction behaviors in a short-video recommendation scenario. Beauty dataset is a subset of the Amazon Review Dataset, which records users’ reviews and interaction behaviors with beauty products on the Amazon platform. FilmTrust dataset contains users’ music listening behaviors. Further analysis of the dataset can be found in the Appendix \ref{sec:dataset}.

\textit{\textbf{Evaluation.}} In this work, we focus on implicit feedback between users and items, where all interactions with ratings greater than 0 are labeled as positive feedback (denoted as 1), and the others as negative feedback (denoted as 0). Following existing work \cite{zhangDualPersonalizationFederated2023, heCoclusteringFederatedRecommender2024}, we adopt the leave-one-out evaluation setting\cite{perifanisFederatedNeuralCollaborative2022} to split these datasets and use Hit Ratio (HR) and Normalized Discounted Cumulative Gain (NDCG) as our evaluation metrics. For fair comparison with existing methods, we calculate Hit@5and NDCG@5 based on the top-5 items in the predicted list.

\subsection{Baselines and Implementation Details}
\textbf{Baselines}
To evaluate the performance of the proposed CGFedRec framework, we compare it with several recent state-of-the-art baselines. To ensure a fair assessment, all models are trained exclusively on implicit feedback. Specifically, this study employs the following methods as baseline methods.
% For detailed configurations and parameter settings of each baseline, see Appendix A.
\begin{itemize}[left=0pt]

\item {\textbf{PFedRec}\cite{zhangDualPersonalizationFederated2023}} proposes a dual personalization mechanism, emphasizing the post-finetuning process to capture personalized information for more accurate user preference modeling.

\item {\textbf{GPFedRec}\cite{zhangGPFedRecGraphGuidedPersonalization2024}} builds a graph neural network-based federated recommendation system that learns the global user-item interaction graph to capture complex collaborative patterns.

\item {\textbf{CoFedRec}\cite{heCoclusteringFederatedRecommender2024}} designs a collaborative FedRec mechanism that leverages inter-user collaborative information to enhance recommendation performance while preserving user data privacy.

\item {\textbf{FedRAP} \cite{li2024federated} introduces additive personalization to integrate globally shared item embeddings with user-specific item embeddings, to improve recommendation performance.}

\item {\textbf{FedCA} \cite{zhangSimilarityPersonalizedFederated2024} proposes a personalized FedRec model with composite aggregation that addresses embedding skew by jointly aggregating similar clients for trained embeddings and complementary clients for non-trained embeddings.}

\end{itemize}

\textit{\textbf{Experimental settings.}} Following the experimental protocol of \cite{he2017neural}, we adopt a 1:4 ratio of positive to negative training pairs. The embedding dimension is set to 32, and the model is optimized using SGD with a batch size of 256. We conduct a grid search over the main hyperparameters: the regularization coefficient $\lambda$ and the temperature parameter $\tau$ are selected from [0.005, 0.01, 0.05, 0.1, 0.5], while the number of clusters $k$ is chosen from [5, 10, 20, 50, 100]. All experiments are implemented in PyTorch and conducted on an NVIDIA GeForce RTX 5090D GPU. 
% The code is publicly available\footnote{https://github.com/buaaRec/CGFedRec.git}.
% The detailed hyperparameter configurations for each dataset are reported in Table \ref{tab:hparam}.
\begin{table*}[]
\renewcommand{\arraystretch}{1.2}
\resizebox{\textwidth}{!}{
\begin{tabular}{lcccccccccc}
\toprule
\multirow{2}{*}{\textbf{Ablation}} 
& \multicolumn{2}{c}{\textbf{ML-100K}} 
& \multicolumn{2}{c}{\textbf{ML-1M}} 
& \multicolumn{2}{c}{\textbf{Filmtrust}} 
& \multicolumn{2}{c}{\textbf{KU}} 
& \multicolumn{2}{c}{\textbf{Beauty}} \\

& \textbf{HR@5} & \textbf{NDCG@5} 
& \textbf{HR@5} & \textbf{NDCG@5}
& \textbf{HR@5} & \textbf{NDCG@5}
& \textbf{HR@5} & \textbf{NDCG@5}
& \textbf{HR@5} & \textbf{NDCG@5} \\
\midrule
\textbf{CGFedRec}               & 0.9777       & 0.8927       & 0.9533      & 0.8851      & 0.9348        & 0.8512        & 0.7402     & 0.5089    & 0.9565       & 0.8458      \\
\textbf{CGFedRec-E}               & 0.5239       & 0.3684       & 0.5475      & 0.3829      & 0.9006        & 0.8170        & 0.3676     & 0.2826    & 0.0632       & 0.0355      \\
\textbf{CGFedRec-EC}               & 0.4836       & 0.3388       & 0.5280       &0.3641             & 0.8981        & 0.8199        & 0.1961     & 0.1219    & 0.0672       & 0.0365      \\
\textbf{CGFedRec-ERC}               & 0.4857       & 0.3405       & 0.5359        &0.3737             & 0.8989        & 0.8215        & 0.2010     & 0.1258    & 0.0672       & 0.0365    \\
\bottomrule
\end{tabular}}
\caption{Ablation results of CGFedRec under different global knowledge sharing strategies. CGFedRec shares global semantic structures by broadcasting only item cluster labels from the server. CGFedRec-EC broadcasts aggregated global item embeddings and cluster labels. CGFedRec-E broadcasts only aggregated global item embeddings without structural signals. CGFedRec-ERC broadcasts aggregated global item embeddings with random cluster labels.}

\label{tab:ablation}
\end{table*}
\subsection{Results \& Discussion (RQ1)}
The experimental results demonstrate that CGFedRec consistently achieves strong recommendation performance across most datasets. This verifies the core assumption of our work: effective federated collaboration does not require strict coordinate-level synchronization of item embeddings. Instead, sharing global semantic structures can provide sufficient collaborative signals to improve local personalization. Compared with embedding aggregation-based baselines, CGFedRec introduces global knowledge in a lightweight form, which avoids forcing clients to fit a unified representation space. This design is particularly beneficial under non-IID interaction distributions, where embedding averaging often produces representation drift and weak semantic consistency.

The improvements on MovieLens and FilmTrust indicate that cluster-level supervision can serve as an effective global structural consistency constraint.. By using cluster labels to construct supervised contrastive signals, CGFedRec pulls semantically related items closer while separating unrelated items. This process stabilizes local representation learning, because the optimization is guided by global relational constraints rather than absolute embedding values. Such constraints are more robust to heterogeneous user preferences, since they encode relative semantic similarity that can be shared across clients. In addition, the server-side clustering mechanism filters redundant embedding details, meaning that only coarse but stable collaborative patterns are retained and broadcast.

Additionally, CGFedRec does not achieve the best performance on the KU dataset. A possible reason is that KU contains fewer users and fewer interactions, which leads to weaker global co-occurrence signals. Under this condition, K-means clustering on aggregated embeddings may become unstable, producing noisy cluster assignments. When cluster labels are unreliable, the contrastive objective may introduce incorrect positive pairs, which weakens representation learning. Overall, the results support that structural alignment is an effective alternative to embedding synchronization.

\subsection{Ablation Study (RQ2)}

We conduct ablation studies to verify the effectiveness of cluster-guided structure alignment and to examine whether embedding-level synchronization is necessary for collaboration. We compare CGFedRec with three variants.

\textbf{\textit{Effect of removing cluster-guided supervision.}} CGFedRec-EC only broadcasts aggregated global item embeddings, without cluster labels, which is equivalent to PFedRec \cite{zhangDualPersonalizationFederated2023}. It only distributes global aggregated embeddings and lacks structural guidance. This approach assumes that embeddings can fully carry global collaborative knowledge. In federated recommendation with non-IID data, directly distributing global embeddings can cause representation drift. Clients may not correctly map the global embeddings to their local preference space, which can lead to conflicts in local optimization. Without clusters to define positive and negative relations, the model cannot effectively capture global semantic structures, which explains its lower performance.

\textbf{\textit{Effect of introducing cluster labels and embedding broadcasting.}} CGFedRec-EC distributes both embeddings and true cluster labels, combining both sources of information. However, its performance does not exceed CGFedRec. The reason is that embedding-level information may conflict with structure-guided constraints. In non-IID settings, the global embeddings introduce bias into local training, while cluster labels impose constraints that require local embeddings to adapt. This conflict can cause gradient interference and reduce convergence quality. 

\textbf{\textit{Effect of introducing random cluster labels and embedding broadcasting.}} Interestingly, CGFedRec-ERC performs slightly better than CGFedRec-EC on most datasets, even though it uses random cluster labels. CGFedRec-ERC distributes global embeddings, but random labels do not impose strict structural constraints. The reason is that clients mainly rely on embeddings for optimization, avoiding gradient conflict caused by real cluster labels. This looser constraint can improve local representation learning and, in some cases, outperform the combination of embeddings and real clusters.
\begin{figure*}
    \centering
    \includegraphics[width=\linewidth]{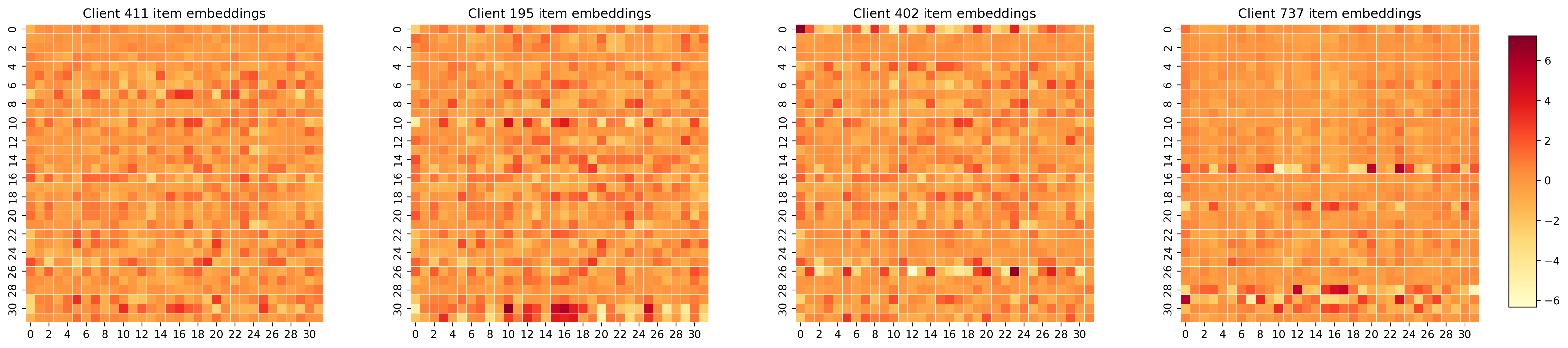}
    \caption{Heatmaps of item embeddings for 4 randomly selected clients in CGFedRec on the ML-100K dataset. Each heatmap visualizes 32 randomly selected items, with color intensity representing embedding values.}
    \label{fig:embeddings}
\end{figure*}

\begin{figure*}
    \centering
    \includegraphics[width=\linewidth]{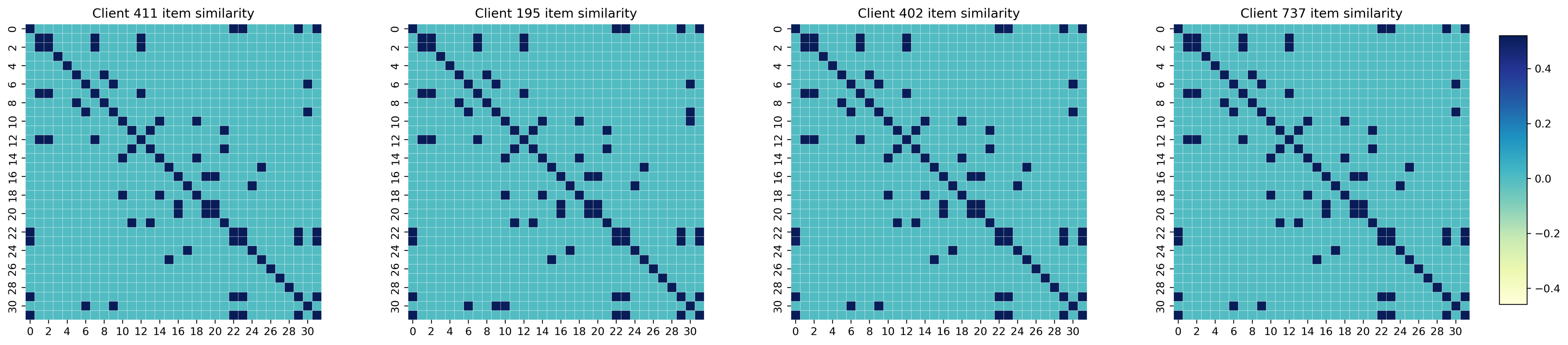}
    \caption{Heatmaps of item-item similarity for 4 selected clients in CGFedRec on the ML-100K dataset. Each heatmap shows cosine similarity computed from item embeddings (Figure \ref{fig:embeddings}) for 32 randomly selected items, with color intensity representing similarity values.}
    \label{fig:similarity}
\end{figure*}

From this analysis, we draw several insights. First, structural signals are a key carrier of global collaborative knowledge in federated recommendation. They are compact, robust, and mitigate the effect of data heterogeneity. Second, directly distributing global embeddings can introduce representation drift and conflict in non-IID settings. Third, combining embeddings and structure requires careful design, it can cause gradient interference and unstable training. Fourth, random cluster labels, although semantically weak, can reduce gradient conflict when used with embeddings, suggesting that the interaction between embeddings and structure is more important than cluster accuracy in some cases.

Overall, the ablation studies confirm the design principle of CGFedRec. Distributing only cluster labels allows efficient communication while maintaining personalized recommendation performance. The results also reveal the interaction between embedding and structure signals in non-IID federated settings, providing insights for designing more robust and efficient structure-guided federated recommendation methods.

\subsection{Personalization and Global Structures Alignment (RQ3)}
To study whether CGFedRec preserves personalization while enforcing global semantic consistency, we analyze local item representations learned on different clients. Figure \ref{fig:embeddings} visualizes item embedding heatmaps of four randomly selected clients. The embedding patterns vary across clients, indicating that item representations are learned in different local coordinate systems. This reflects the personalized nature of local training, where each client forms its own feature space based on its interaction distribution.

To examine whether semantic structure is still aligned, we compute item-item cosine similarity matrices induced by local embeddings, as shown in Figure \ref{fig:similarity}. Although embeddings are not coordinate-aligned, the similarity matrices show consistent block structures across clients. Items belonging to the same semantic group form stable similarity clusters, suggesting that local models preserve consistent relative relations among items.

This observation matches the design principle of CGFedRec. Each client can learn personalized embedding coordinates while producing a relational geometry that follows the same semantic manifold. Therefore, CGFedRec achieves global semantic alignment through structure consistency, rather than coordinate consistency. 

Overall, the results demonstrate that CGFedRec preserves local personalized embedding spaces while injecting global semantic constraints, which redefine the concept of federated collaboration. Additional t-SNE visualizations are provided in Appendix \ref{sec:vis}.

\begin{figure}
    \centering
    \includegraphics[width=\linewidth]{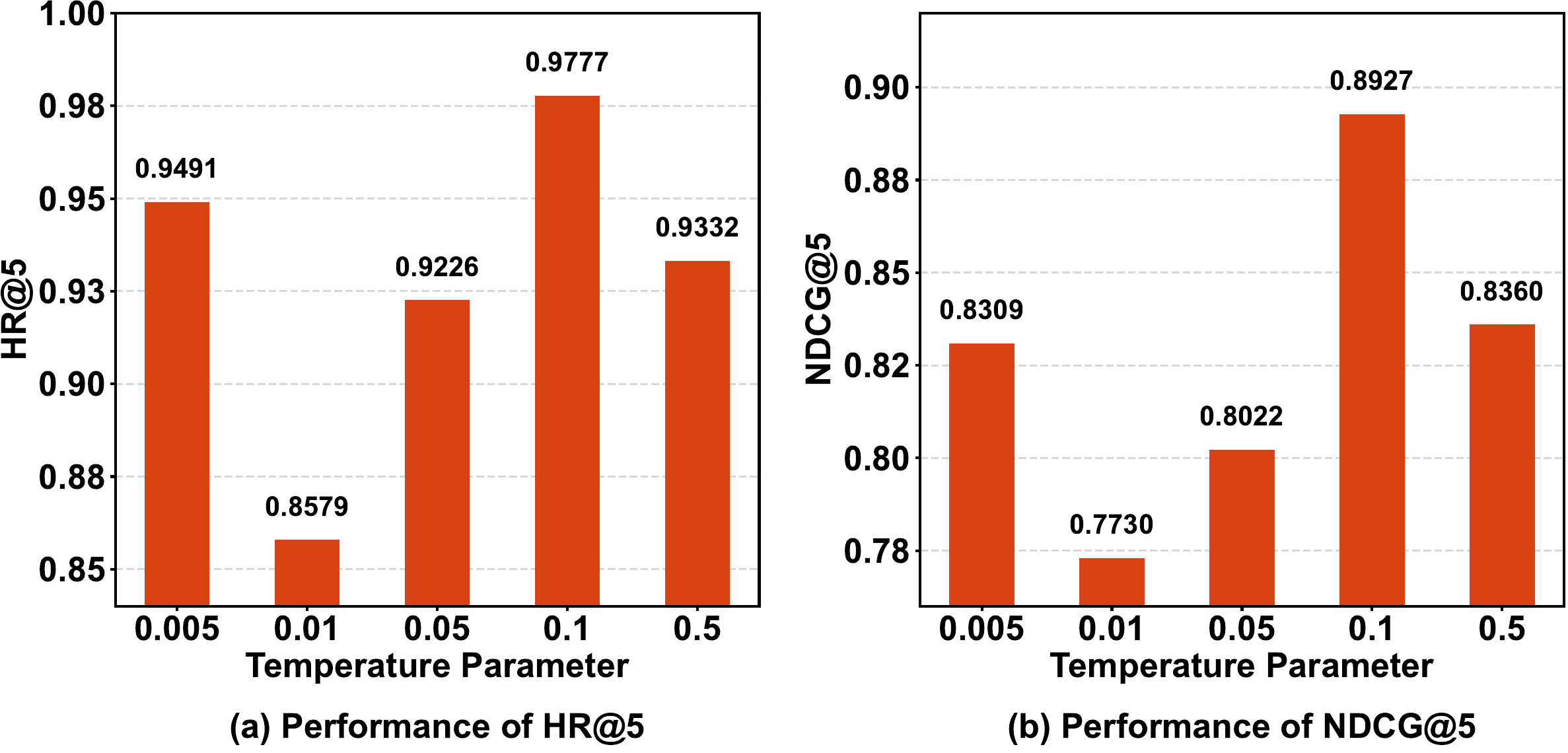}
    \caption{Effect of the temperature parameter on the performance of CGFedRec on the ML-100K dataset.}
    \label{fig:t}
\end{figure}

\subsection{Parameter Sensitivity Analysis (RQ4)}
In this section, we analyze the impact of three key hyper-parameters in CGFedRec, including the temperature parameter $\tau$ in the cluster-guided contrastive learning, the number of clusters $K$ used for global structure discovery, and the regularization coefficient $\lambda$ that balances $\mathcal{L}_{cg}$ with the main recommendation objective.

\textbf{\textit{Effect of Temperature Parameter.}} We investigate the influence of the temperature parameter $\tau$, which controls the sharpness of the contrastive distribution. As shown in Figure \ref{fig:t}, the best performance is achieved when $\tau=0.1$. When $\tau$ is too large, the softmax distribution becomes smooth, which reduces the discrimination between positive and negative pairs. Thus, the cluster-based structural signal cannot effectively separate item representations. When $\tau$ is too small, the distribution becomes extremely sharp, which amplifies hard negatives and increases optimization instability. Since cluster labels are coarse-grained supervision, overly strict alignment may introduce noise and reduce generalization. These results indicate that moderate $\tau$ enables effective structure injection while maintaining stable training.

\begin{figure}
    \centering
    \includegraphics[width=\linewidth]{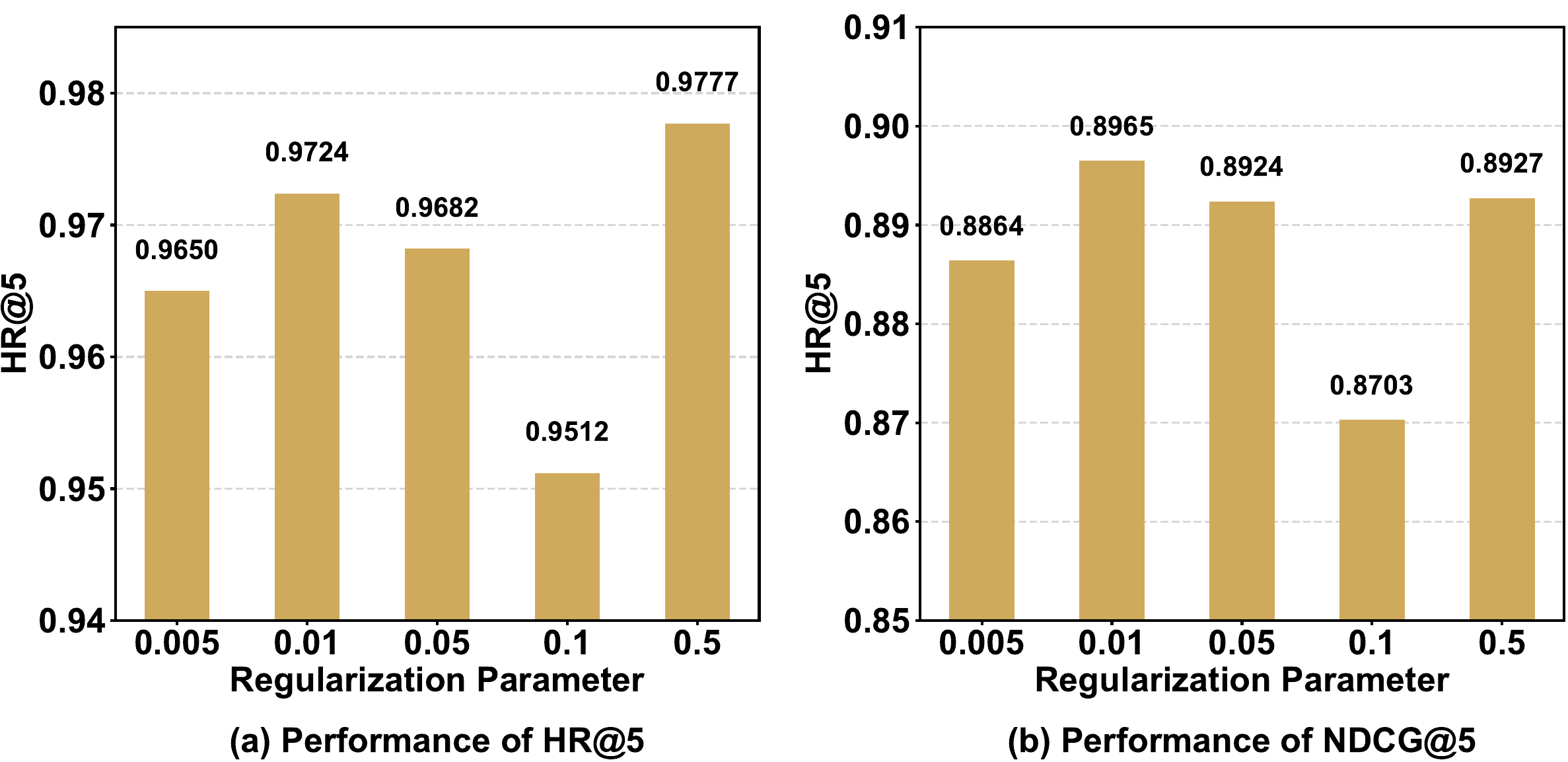}
    \caption{Effect of the coefficient of regularization on the performance of CGFedRec on the ML-100K dataset.}
    \label{fig:reg}
\end{figure}

\textit{\textbf{Effect of Regularization Coefficient.}} We then analyze the regularization coefficient $\lambda$ that weights the contrastive loss $\mathcal{L}_{cg}$. As shown in Figure \ref{fig:reg}. When $\lambda$ is too small, $\mathcal{L}_{cg}$ becomes negligible, and the model mainly relies on local training, limiting global collaboration. When $\lambda$ is too large, training is dominated by structural alignment, which strengthens global consistency and improves recall, leading to better HR. However, it may suppress personalization signals, which are important for fine-grained ranking, thus limiting NDCG. A moderate $\lambda$ provides a better balance, enabling structure guidance while preserving local ranking flexibility.

\begin{figure}
    \centering
    \includegraphics[width=\linewidth]{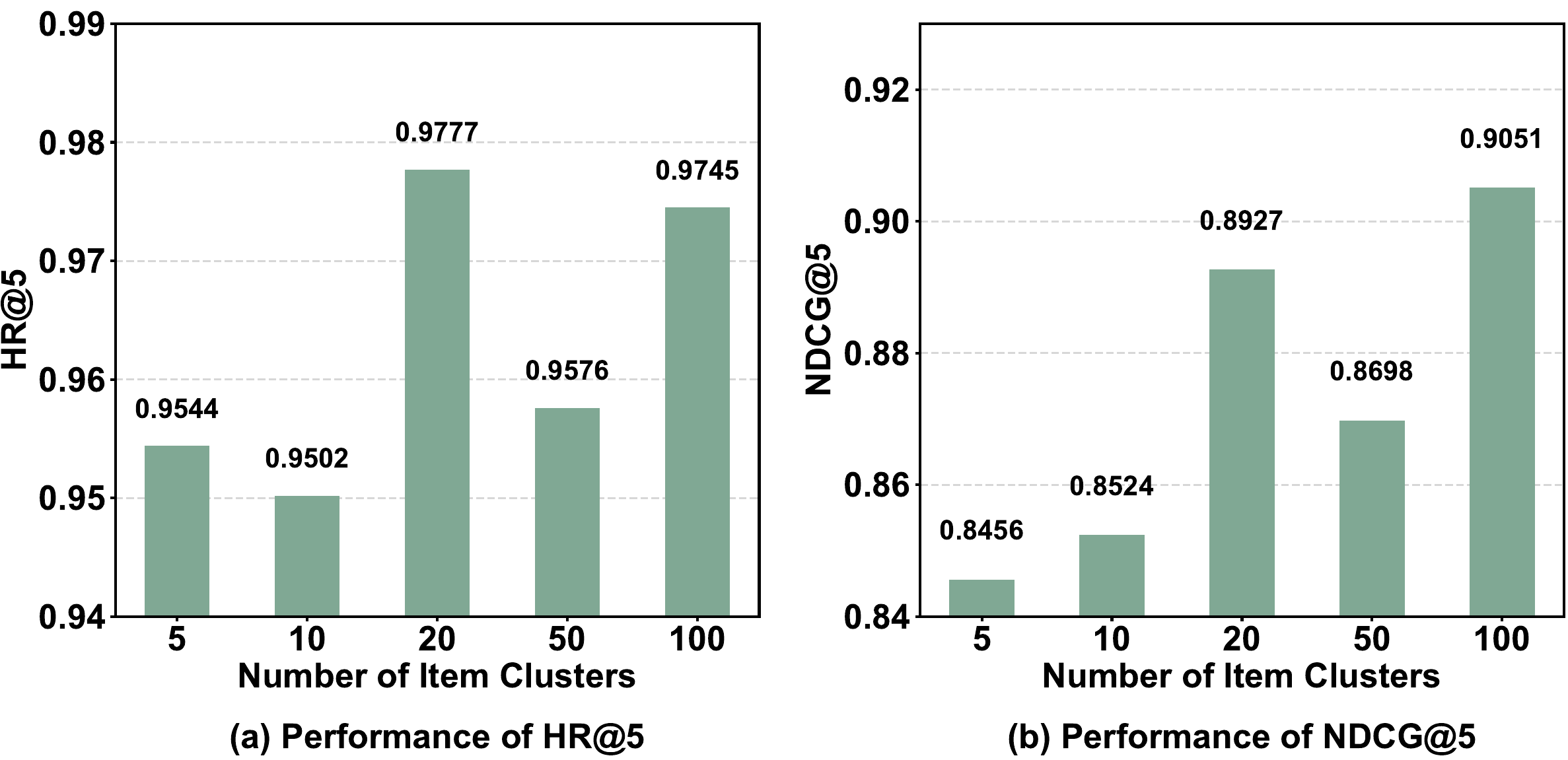}
    \caption{Effect of the number of clusters on the performance of CGFedRec on the ML-100K dataset.}
    \label{fig:k}
\end{figure}

\textit{\textbf{Effect of Cluster Number.}} We further study the number of clusters $K$, which determines the granularity of the global semantic structure. As shown in Figure \ref{fig:k}. When $K$ is small, semantically different items are grouped into the same cluster, introducing false positives in contrastive learning. This weakens representation discrimination and harms ranking quality. When $K$ increases, clusters become more informative and reduce noisy positive pairs, improving both metrics. When $K$ is large, each cluster contains fewer items, which reduces the number of positive pairs and weakens the contrastive signal. However, fine-grained clusters provide more precise semantic separation, which benefits top-ranked ordering.

Overall, the results suggest that CGFedRec requires an appropriate calibration between global structure supervision and local personalization learning.

\subsection{Ablation Study on Differential Privacy Noise}

\begin{table}[htbp]
\begin{tabular}{cllllll}
\toprule
 \textbf{Intensity} $\delta$     & \multicolumn{1}{c}{\textbf{0}} & \multicolumn{1}{c}{\textbf{0.1}} & \multicolumn{1}{c}{\textbf{0.2}} & \multicolumn{1}{c}{\textbf{0.3}} & \multicolumn{1}{c}{\textbf{0.4}} & \multicolumn{1}{c}{\textbf{0.5}} \\
 \midrule
\textbf{HR@5}  & 0.9777                & 0.9936                  & 0.9756                  & 0.9724                  & 0.9894                  & 0.9788                  \\
\textbf{NDCG@5} & 0.8927                & 0.9294                  & 0.8954                  & 0.8731                  & 0.9086                  & 0.8894   \\
\bottomrule
\end{tabular}
\caption{Results of applying local differential privacy into CGFedRec with various noise intensity on the ML-100K.}
\label{tab:dp}
\end{table}
% \textcolor{red}{compact}
% We observe that moderate differential privacy noise improves recommendation performance over non-private baselines. 
% This counter-intuitive phenomenon stems from the regularization effect of benign noise on manifold learning~\cite{wang2016learning,noh2017regularizing}. 
% Client embeddings often carry high-frequency noise due to local overfitting. 
% Injected noise acts as a smoothing filter on the embedding space. 
% It masks non-essential local variations and prevents the server from fitting client-specific biases. 
% Furthermore, random perturbations mitigate the impact of outliers on clustering centroids. 
% This forces cluster assignments to capture the underlying probability density of the item manifold. 
% Thus, privacy noise serves as a structural regularizer. 
% It facilitates convergence to flatter and more generalizable minima.

% \textcolor{red}
In this subsection, we evaluate the performance of the proposed privacy-enhanced CGFedRec under the local differential privacy (LDP) strategy. Specifically, we vary the noise intensity parameter $\delta$ over the range $[0, 0.1, 0.2, 0.3, 0.4, 0.5]$, and the experimental results are reported in Table \ref{tab:dp}. 

Contrary to the common assumption that privacy comes at the cost of utility, our experiments reveal a counter-intuitive phenomenon. The model with moderate noise injection consistently outperforms the non-private baseline. This observation can be attributed to the regularization effect of benign noise on the manifold learning process~\cite{wang2016learning,noh2017regularizing}. Client-side embeddings often contain high-frequency noise derived from local overfitting to sparse interactions. When uploaded directly, these fluctuations distort the global geometric structure. The injected differential privacy noise effectively acts as a smoothing filter on the embedding space. 
It masks these non-essential local geometric variations. Consequently, the server extracts more robust global semantic structures rather than fitting to client-specific biases. Furthermore, random perturbations enhance the stability of cluster centroids. Noise prevents the clustering algorithm from being skewed by outlier embeddings. This forces the cluster assignments to capture the underlying probability density of the item manifold. Therefore, moderate privacy noise serves as a structural regularizer in CGFedRec. It prevents the memorization of noise patterns and helps the optimization process converge to flatter and more generalizable minima. This results in a synergy where privacy protection and recommendation accuracy are simultaneously improved.

\section{Conclusion}
In this paper, we revisit the fundamental collaboration mechanism in federated recommendation and argue that effective global knowledge sharing does not necessarily require strict embedding-level synchronization. Instead, maintaining consistent semantic structures across items can serve as a more compact and robust carrier of collaborative signals. Based on this structural perspective, we propose CGFedRec, a cluster-guided federated recommendation framework that learns global item clusters on the server and broadcasts only lightweight cluster labels to clients. By employing these labels to construct cluster-based contrastive supervision, CGFedRec enables clients to align local item representations under shared semantic constraints while preserving personalization flexibility. This design explicitly eliminates downstream transmission of high-dimensional item embeddings, reducing communication complexity from $O(nd)$ to $O(n)$. Extensive experiments on multiple real-world datasets demonstrate that CGFedRec consistently achieves superior recommendation accuracy compared with state-of-the-art federated recommendation baselines, while significantly improving communication efficiency. Our results highlight the potential of structural collaboration as a promising direction for scalable and practical federated recommendation systems.

%Bibliography
\bibliographystyle{unsrt}  
\bibliography{ref}  

\section{Appendices}
\begin{figure*}
    \centering
    \includegraphics[width=\linewidth]{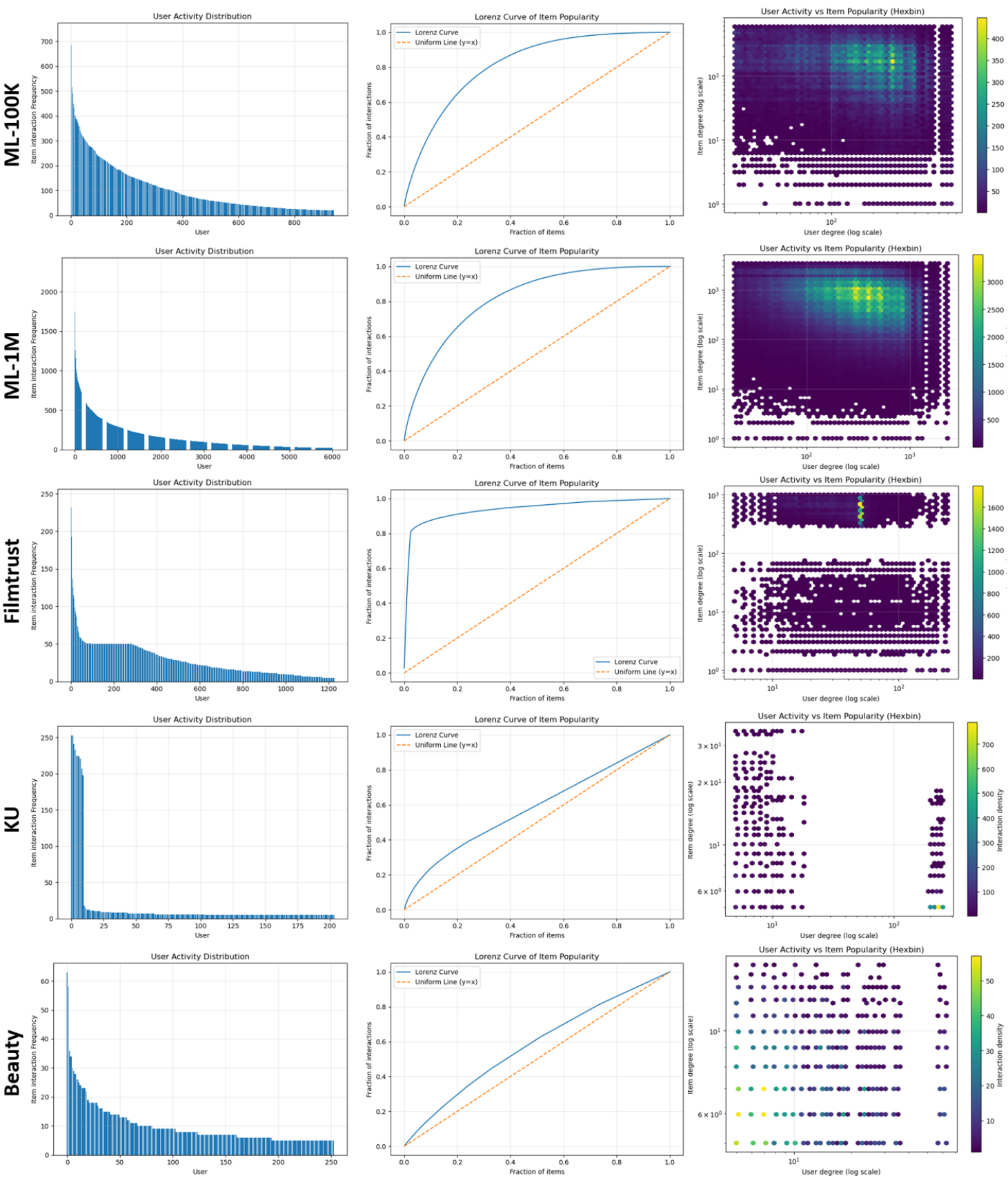}
    \caption{Statistical characteristics and interaction distributions of the five benchmark datasets. The columns from left to right represent: (a) User Activity Distribution, showing the long-tail nature of user interactions; (b) Lorenz Curve of Item Popularity, illustrating the concentration of interactions among items (the dashed line represents perfect equality); (c) Hexbin Density Plots of User-Item Degrees, visualizing the correlation between user activity and item popularity on a logarithmic scale.}
    \label{fig:datasetAna}
\end{figure*}
\begin{figure*}
    \centering
    \includegraphics[width=\linewidth]{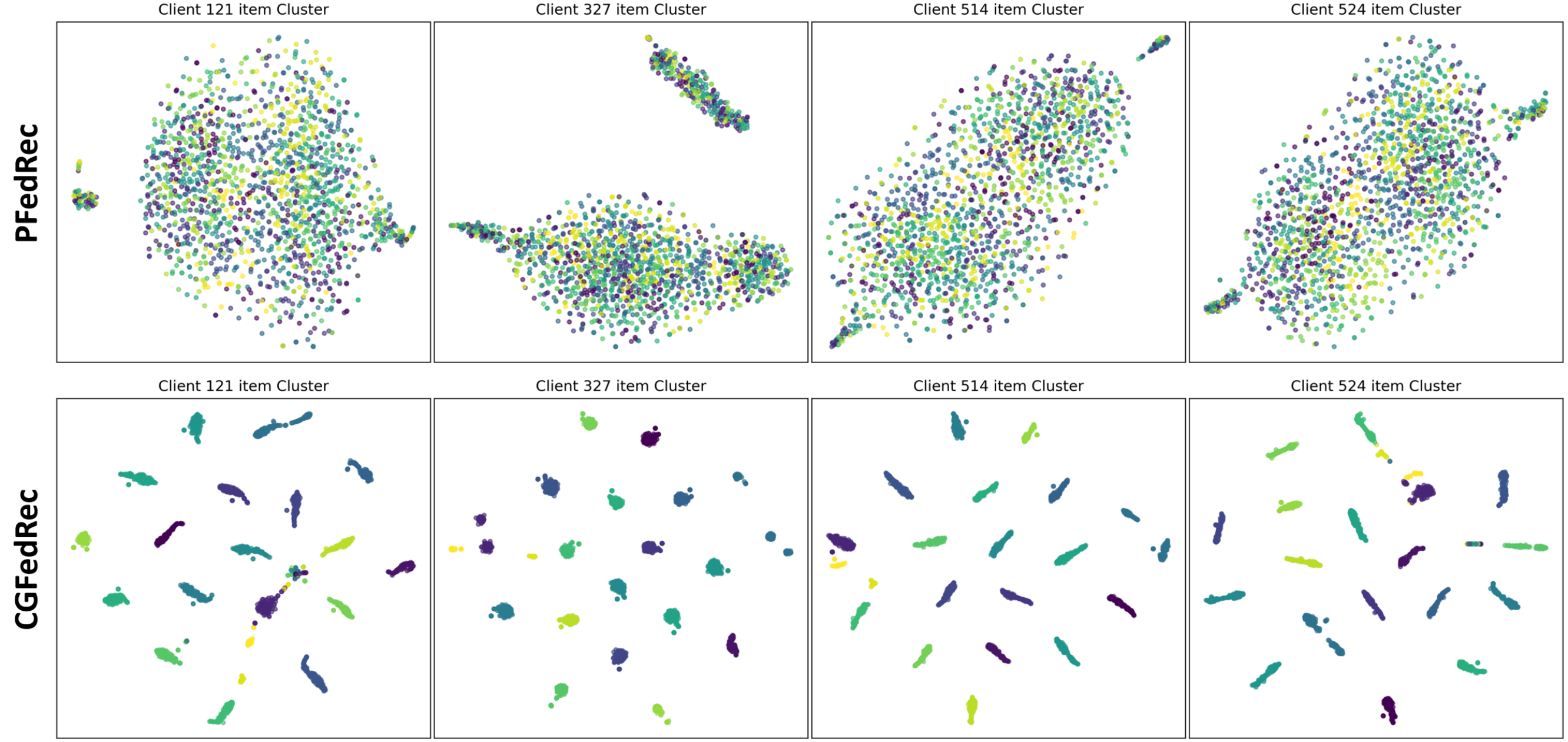}
    \caption{T-SNE visualization comparing item clustering representations of PFedRec and CGFedRec.}
    \label{fig:cluster}
\end{figure*}
\subsection{Dataset Characteristics and Exploratory Analysis}\label{sec:dataset}
To characterize the underlying patterns of user-item interactions and validate the robustness of our model across diverse scenarios, we conduct an exploratory data analysis on five benchmark datasets: MovieLens-100K (ML-100K), MovieLens-1M (ML-1M), FilmTrust, KU, and Beauty. Figure \ref{fig:datasetAna} illustrates the statistical distributions through three distinct lenses: user activity, item popularity concentration, and the correlation between user-item degrees.

\textit{\textbf{User Activity and Interaction Frequency.}} The first column of the Figure presents the User Activity Distribution, where users are ranked by their interaction frequency. All datasets demonstrate a consistent long-tailed distribution, though the head of the distribution varies significantly. In ML-1M and ML-100K, a substantial number of active users contribute hundreds of interactions, whereas in the KU and Beauty datasets, the frequency drops sharply after the first few dozen users. This variation in user engagement profiles allows us to evaluate how effectively the model captures preferences for both "heavy" users and those with extremely limited interaction history.

\textit{\textbf{Concentration of Item Popularity.}} The second column displays the Lorenz Curve of Item Popularity, which plots the cumulative fraction of interactions against the cumulative fraction of items. The deviation from the diagonal "Uniform Line" ($y=x$) quantifies the degree of popularity bias within the systems. Notably, the FilmTrust dataset exhibits the most extreme curvature, indicating that a very small percentage of items accounts for the vast majority of interactions. Conversely, the Beauty and KU datasets show curves closer to the diagonal, suggesting a more decentralized interaction pattern. Modeling these differing levels of concentration is critical, as it tests the ability of a recommender system to alleviate popular-item bias and surface long-tail content.

\textit{\textbf{Correlation between User and Item Degrees.}} The third column utilizes Hexbin density plots to visualize the relationship between user degree and item degree (both on a log scale). These plots reveal the structural density of the interaction space. ML-100K and ML-1M show high-density clusters in the mid-to-high degree regions, indicating that active users frequently interact with popular items. FilmTrust exhibits a distinct stratified structure, suggesting a highly partitioned interaction matrix where specific groups of users are concentrated around certain popularity tiers. KU and Beauty datasets show scattered, low-density distributions due to their high sparsity and smaller user/item pools.

These visualizations highlight two primary challenges for recommendation models. First, the imbalance in user activity requires models to generalize from sparse signals for inactive users. Second, the Lorenz Curve deviations confirm that the datasets are heavily influenced by popularity bias, necessitating mechanisms that can distinguish between genuine user interest and global item trends. By evaluating our proposed method across these varied statistical landscapes, we ensure its effectiveness in handling both dense, popular-driven environments and sparse, niche-oriented scenarios.

\subsection{Visualization of clustering effects.}\label{sec:vis}

Figure \ref{fig:cluster} presents the t-SNE visualization of item embedding distributions on four randomly selected clients. We compare PFedRec and our CGFedRec under the same clustering pipeline, where the server computes global item embeddings by averaging uploaded client embeddings and then performs K-means clustering to obtain global cluster assignments. The visualization is conducted based on each client’s local item embeddings, with points colored by the global cluster labels.

For PFedRec, the item embeddings on different clients exhibit weak cluster separability. Most items are distributed in a highly mixed manner, where points with different labels overlap heavily. This phenomenon indicates that embedding-level synchronization does not guarantee semantic alignment across clients. Under heterogeneous interaction distributions, the aggregated global embeddings tend to be biased toward dominant clients, which leads to inconsistent geometric structures on individual clients. As a result, the same global cluster label may correspond to scattered regions in the local embedding space, implying that PFedRec fails to enforce stable semantic grouping.

In contrast, CGFedRec produces clearer cluster-aware structures across all clients. Items sharing the same global label form compact local groups, while items from different clusters are pushed into separated regions. This pattern is consistent with the proposed cluster-guided contrastive objective, which explicitly treats intra-cluster item pairs as positives and inter-cluster pairs as negatives. The visualization suggests that CGFedRec enables clients to preserve global semantic relations without requiring coordinate-level embedding consistency. Meanwhile, local embedding flexibility is maintained, since each client can still learn personalized representations within the cluster-level constraints.

Overall, the results verify that cluster labels serve as an effective structural carrier of collaborative signals. They guide local representation learning in a way that improves semantic coherence and mitigates heterogeneity-induced representation drift, while avoiding downstream embedding transmission.
\end{document}